\documentclass[twocolumn,prd,superscriptaddress,showpacs,floatfix,%
preprintnumbers,nofootinbib]{revtex4}
\usepackage{graphics}
\usepackage{bm}
\usepackage{epsfig}
\usepackage{graphicx}
\usepackage{amsmath}

\newcommand{\gtrsim}{\,\rlap{\lower3.7pt\hbox{$\mathchar\sim$}}
\raise1pt\hbox{$>$}\,}
\newcommand{\lesssim}{\,\rlap{\lower3.7pt\hbox{$\mathchar\sim$}}
\raise1pt\hbox{$<$}\,}

\begin{document}

\title{Dark Matter Decaying into a Fermi Sea of Neutrinos}
\author{Ole Eggers Bj\ae lde}
\affiliation{Institut f\"ur Theoretische Teilchenphysik und
Kosmologie RWTH Aachen University, D - 52056 Aachen, Germany,
bjaelde@physik.rwth-aachen.de} \affiliation{Department of Physics
and Astronomy, University of Aarhus, Ny Munkegade, Bld. 1520,
DK-8000 Aarhus\, C}
\author{Subinoy Das}
\affiliation{Department of Physics and Astronomy, University of British Columbia, BC, V6T 1Z1 Canada, subinoy@phas.ubc.ca}

\preprint{TTK-10-15}

\date{{\today}}

\begin{abstract}
We study the possible decay of a coherently oscillating scalar
field, interpreted as dark matter, into light fermions.
Specifically, we consider a scalar field with sub-eV mass decaying
into a Fermi sea of neutrinos. We recognize the similarity between
our scenario and inflationary preheating where a coherently
oscillating scalar field decays into standard model particles. Like
the case of fermionic preheating, we find that Pauli blocking
controls the dark matter decay into the neutrino sea. The radius of
the Fermi sphere depends on the expansion of the universe leading to
a time varying equation of state of dark matter. This makes the
scenario very rich and we show that the decay rate might be
different at different cosmological epochs. We categorize this in
two interesting regimes and then study the cosmological
perturbations to find the impact on structure formation. We find
that the decay may help alleviating some of the standard problems
related to cold dark matter.
\end{abstract}
\pacs{95.35.+d, 98.80.-k}

\maketitle

%%%%%%%%%%%%%%%%%%%%%%%%%%%%%%%%%%%%%%%%%%%%%%%%%%%%%%%%%%%%%%%%%%%%%%
\section{Introduction}

Dark matter has become an extremely interesting area of research in
both cosmology and particle physics. From the particle physics point
of view it can be thermal WIMPs, axions, Kaluza Klein states, etc.
Though supersymmetric (SUSY) models predict its mass to be of the
order the electro-weak scale, there are viable models of dark matter
where its mass can be as low as sub-eV, for example axion-like dark
matter. Especially the direct and indirect search for dark matter
has narrowed down the parameter space of these well studied
candidates to a large extent. Hence it is highly likely that dark
matter is of much more exotic nature than thought of. In addition,
there may be a requirement for more complicated physics such as
interactions with dark energy or with neutrinos. Similarity between
the neutrino mass and the present dark energy density scale has
inspired people to look for a connection between the two. Now we know
that the normal active neutrino cannot be a viable dark matter
candidate because of its free-streaming ability. But the existence
of sub-eV neutrino mass might point towards richer sub-eV scale
physics. In fact there has been a few interesting works
~\cite{Fardon:2005wc, Chacko:2004ky} where new states of sub-eV
masses are present in the dark sector. The reason is, if the dark
sector interacts only gravitationally with the standard model
sector, a TeV scale SUSY breaking in SM would predict scalar
particles of mass $\frac{TeV^2}{M_{Pl}} \sim 10^{-3}$eV in the dark
sector. Scalar dark matter of milli eV mass with a possible coupling
to neutrinos has been discussed in ~\cite{Das:2006ht}. Also moduli
of sub-eV mass can easily arise from string compactification
\cite{Kaloper:2008qs}.

If in nature dark matter arises from such a low energy scale, we
would expect it to decay into light fermions like neutrinos through
a Yukawa type of coupling. GeV scale dark matter decay (and
annihilation) to neutrinos has drawn lots of recent interest
\cite{Mardon:2009rc, Buckley:2009kw,
Spolyar:2009kx,Allahverdi:2009se,Hisano:2008ah,Liu:2009ac,Covi:2009xn},
especially in the context of recent cosmic ray measurements and the
DAMA/LIBRA experiment \cite{Hooper:2008cf,Kumar:2009ws}. But in our
case, dark matter is of sub-eV mass and the signature of its decay
into neutrinos is mainly cosmological, especially in structure
formation. Recently, in a different context, \textit{the non-thermal
wimp miracle} \cite{Acharya:2009zt} was introduced where a scalar of
TeV mass decays into a stable dark matter particle. So the decay of a
scalar particle can lead to very rich  phenomenology in cosmological
context.

The possible decay of scalar dark matter into light neutrinos is an
effect which could potentially help to understand the apparent
surplus of power on small scales in simulations containing normal
CDM \cite{Navarro:1996gj,Klypin:1999uc,Moore:1999nt}. This surplus
could for instance be an indication that CDM is simply clustering
too much on small scales and we need some mechanism to reduce their
gravitational interaction. This is where the decay could play a role
- see \cite{Sahni:1999qe} for a similar idea. In this paper we study
the nature of the decay and its possible signature in structure
formation. As we consider an axion-like scalar dark matter, it is
undergoing a coherent oscillation and might decay into neutrinos
through a parametric excitation. In that case, the process will have
many similarities with inflationary fermionic preheating
\cite{Greene:1998nh,Giudice:1999fb,GarciaBellido:2000dc,Greene:2000ew}.
Therefore we dub the process \textit{Preheating Dark Matter}.

Cosmology with decaying/interacting dark matter has been an
interesting topic of research
\cite{Das:2005yj,Guo:2007zk,Olivares:2007rt,Boehmer:2008av,Ibarra:2008jk,Cirelli:2009dv,Ibarra:2009dr}
in recent times as it gives a probe to detect dark matter indirectly
through its effects on structure formation. On top of that, if it
decays into dark energy it might give rise to a unified description
of dark matter and dark energy. In most of the studies
\cite{Valiviita:2008iv,Cen:2000xv}, the rate of energy transfer from
dark matter to other components (eg. dark energy, radiation or
neutrinos) has been empirically assumed driven by mathematical
simplicity. Here we present a concrete model of dark matter decay to
light fermionic states like neutrinos and then study its imprint on
structure formation. In particular, we derive the decay rate of dark
matter as a function of redshift using the theory of fermionic
preheating. We find that in the initial stage the decay rate is
faster and determined by parametric resonance. But at late times the
parametric production ceases and redshifts of fermionic modes
control the decay rate. This time varying decay rate makes the
phenomenology rich and offers a prominent imprint on the structure
formation, something which might be experimentally probed by near
future experiments.

The plan of the paper is as follows: In section II we discuss the
particle physics aspect of our scenario. In section III, we
incorporate the idea of inflationary preheating for scalar dark
matter decaying into neutrinos. In section IV, we identify different
epochs of decay and derive the background evolution, i.e. how dark
matter and neutrino energy densities evolve in presence of the
decay. In section V, we derive the perturbation equations for our
scenario and in section VI we obtain temperature anisotropy and
matter power spectra. Finally, we summarize our work in section VII.

\section{Phenomenological Model}
Here we discuss how our scenario fits into a particle physics set
up. Scalar fields of sub-eV masses are common in different particle
physics models. In many models of TeV scale gauge mediated supersymmetry
breaking, a gravitationally coupled dark sector contains a
scalar of sub-eV mass. Also, string compactifications generically
predict axion like scalars such as the dilaton and large numbers of
moduli \cite{Jaeckel:2010ni,Piazza:2010ye} whose mass can easily be
in the sub-eV range.

As we are interested in a sub-eV scalar field which couples to light
fermions like the standard model neutrino, our set up is inspired by
the models of mass varying neutrino dark energy \cite{Fardon:2003eh}
where a sub-eV mass scalar couples to the standard model neutrino -
see also
\cite{Takahashi:2005kw,Bjaelde:2007ki,Bean:2007ny,Antusch:2008hj,Bjaelde:2008yd}.
Especially, we refer to a model of supersymmetric neutrino dark
energy, where multiple scalars of such low mass are present and can
couple to neutrinos. In such a theory, it has been shown that the
scalar potential takes the form of the well known hybrid inflation
potential. We refer readers to \cite{Fardon:2005wc} for details and
will briefly discuss here. The Lagrangian for mass varying neutrino
dark energy is given by
\begin{equation}
{\cal L} \supset m_D \nu N + \kappa A N N + h.c. + V(A)
\end{equation}
where $m_D$ is the Dirac mass and $\nu$ is the left chiral Weyl
field representing active neutrinos. N is the righthanded heavy
fermion and A is the scalar. $\kappa$ is some Yukawa coupling.

In supersymmetric models of neutrino dark energy where $ \nu, N, A$
is promoted to superfields $ l, n, a$, the superpotential takes the
form
\begin{equation}
W = \kappa a n n + m_D l n
\end{equation}

After taking quantum corrections into account, it has been shown
that this leads to a hybrid inflation kind of potential. Depending on
the temperature of the universe, the scalar either remains trapped
at a metastable minimum playing the role of dark energy or it rolls
off and starts oscillating coherently, behaving like cold dark
matter. Following a simple model \cite{Das:2006ht}, the Lagrangian
looks like
\begin{align}
{\cal L}=&\lambda n_2 \psi_3^2+ 2 \lambda n_2  \psi_2 \psi_3+m_3
\psi_3 \nu_3+m_2 \psi_2
\nu_2\nonumber\\
+&V_{susy}+V_{soft}+V_{\epsilon}+h.c.
\end{align}
where
\begin{equation}
V_{susy} = 4 \lambda^2 n_2^2 n_3^2 + \lambda^2 n_3^4,
\end{equation} and
\begin{equation}
V_{soft} = \tilde m_2^2 n_2^2-\tilde m_3^2 n_3^2+ \tilde a_3 n_3^3
\end{equation}
The terms in $V_\epsilon$ are included in order to generate a
Majorana mass for the neutrino in the vacuum. In this kind of theory
the superpartner sneutrinos (here denoted by $n_2, n_3$) can easily
be of sub-eV mass and play the role of dark matter. Also it can
easily couple to light fermions (neutrinos). From now on, we will
switch our focus to cosmological effects of such a model.

\section{Preheating from scalar dark matter}
We are essentially interested in a light scalar field dark matter of
sub eV mass which has a coupling to an ultra light fermion which for
our case is the standard model neutrino. We consider decays of such
dark matter into neutrinos, though our framework is true for decay
into any fermion. Following the mechanism of inflationary
preheating, in this section we will understand the physical nature
of the decay and will clarify different regimes of decay. We will
see that the decay rate changes as the universe expands due to time
evolution of a resonance parameter which controls the parametric
excitation of neutrinos.

We mainly follow the work ~\cite{Greene:2000ew} on fermionic
preheating and apply that to our scenario. So we refer to this work
for detailed derivation of the equations. Briefly, to find the
number density of created fermions through the preheating mechanism
in an expanding background, one derives a mode equation using the
original Dirac equation with the Friedmann-Lema\^{\i}tre-Robertson-Walker
metric. It has been shown that the comoving number density of
created fermions can be obtained by solving for a mode function
$X_k(t)$. For a Yukawa type coupling $\lambda\phi \psi \bar{\psi}$,
the mode equation is given by
\begin{equation}
X_k^{''} + [\kappa^2 + (\tilde{m} + \sqrt{q} f)^2 -i \sqrt{q} f']
X_k =0,
\end{equation}
where $\phi_{0}f(t)$ is the background solution for the time
evolution of the oscillating scalar field, $\kappa \equiv
\frac{k}{m_{\phi}}$ is the dimensionless fermion mass, $\tilde{m}
\equiv \frac{m_{\psi}}{m_{\phi}} $, and the resonance parameter  $q
\equiv \frac{\lambda^2 \phi_0^2}{m_{\phi}^2}$. These three
parameters completely determine the parametric production of
fermions. We consider the oscillation of the field with the usual
quadratic potential $V= \frac{1}{2} m^2 \phi^2 $ as this is a good
approximation around minima of any potential. The term $(\tilde{m} +
\sqrt{q} f)$ can be thought of as an effective mass of the fermion.
As the scalar field oscillates, the effective mass itself will
oscillate around zero and the parametric production of fermions is
enhanced when the effective mass crosses zero. It has been shown
numerically that $n_k(t)$ oscillates and due to Pauli blocking its
maximum value never crosses unity. But for decay into bosonic
particles it is not bounded by unity. We stress that the behavior of
this parametric production is considerably different than the
perturbative decay process $\phi \rightarrow \bar{\psi} \psi $ where
the decay rate is given by $\Gamma \simeq \frac{\lambda^2 m}{8
\pi}$.

In the above mode equation, expansion of the universe has been
neglected which may only be true at very late times where the Hubble
parameter drops. To get a full understanding, one must include the
expansion of the universe. This alters two aspects. The parameters $q$
and $\kappa$ now become time dependent. More specifically we get
$q\equiv \frac{\lambda^2 \phi^2(\tau)}{m_{\phi}^2}$ and the physical
momentum $p \equiv \frac{\kappa}{a(t) m_{\phi}}$ where $a(t)$ is the
scale factor of the universe. As a result, the periodic modulation
of the comoving number density does not hold anymore. For large
values of the resonance parameter $(q \geq 1)$, the calculation of
parametric production becomes, in fact, simple. Luckily, we will see
later that for our case $q \gg 1$ for large periods of
(cosmological) time. Using the method of successive scattering for
fermions, it has been shown \cite{Greene:2000ew} that due to the
loss of periodicity of $n_k$ the production of fermions happens
through a stochastic filling of a Fermi sphere up to a Fermi radius
$\kappa_F$ which depends on scale factor $a(t)$ and is given by

\begin{equation}
\kappa_F^2 \simeq \sqrt{q(t)^{1/2}} \, a(t).
\end{equation}

Now to find the exact dependence, one needs to know how $q(t)$
changes with scale factor. As we are using a quadratic potential for
the scalar, the solution for the scalar field for this case is well
known. Oscillation of $\phi$ in this case is given by the asymptotic
solution
\begin{equation}
 \phi (t) \sim \frac{\phi_0}{a^{3/2}} \cos(t).
\end{equation}
Using this it is easy to derive
\begin{equation}
 \kappa_F = m_{\phi} \, q_0^{1/4} \, a^{1/4},
\end{equation}
where $q_0 \equiv \frac{\lambda^2 \phi_0^2}{m_{\phi}^2}$.

So, as the universe expands, the Fermi sphere also expands producing
more and more neutrinos. But the resonance parameter decreases as
the amplitude of oscillation drops due to Hubble friction and at
some point the Fermi sphere stops expanding when  $q(t)$ becomes of
the order of unity. At this regime, the redshift of fermionic modes
due to Hubble expansion is fast enough to prevent the parametric
excitation. Finally, fermions will be produced with a much lower rate
in the perturbative regime and perturbative processes continue
unless $ m_{\phi} < 2 m_{\psi}$.

\section{Different decay regimes}

Using the above results now we can focus on production of neutrinos
and its time evolution. Here we assume that parametrically produced
neutrinos mix with other relic neutrinos and acquire the same
temperature through thermalization. It is instructive to note that
as the mass of the scalar is way less than in the usual inflationary
preheating scenario, we would get parametric excitation until very
late times. From the previous discussion, we have learnt that the
resonance parameter $q$ is very crucial to determine the nature of
the decay and $q$ itself is time dependent. Now we will discuss the
two different decay regimes and the transition time between them for
our simple model with a quadratic potential.

\subsection{Regime I: Expanding Fermi radius  ($q>>1$)}

During early stages of parametric production
\begin{equation}
 q(t) \equiv \frac{\lambda^2 \phi(t)^2 }{m_{\phi}^2} =  2 \, \lambda^2 \frac{\rho_{\rm
 DM}}{m_{\phi}^4}.
\end{equation}
As we are interested in scalar mass of the order of $ m_{\phi} \sim
10^{-3}$ eV, almost all over the cosmic history until today,
$\rho_{\rm DM} \geq (10^{-3} eV)^4 $. Now if the coupling constant
$\lambda$ is of the order of unity, we still get parametric
production at very late times. But for smaller couplings parametric
excitation stops at earlier redshift when $q(t) \simeq 1$. Later we
will take different choices of the coupling $\lambda$ and study how it
affects the formation of structure. The produced neutrino number
density is obtained through the volume of the Fermi sphere with
radius
\begin{equation}
 \kappa_F^{\rm phys} = q^{1/4} a^{1/4} \times a^{-1} = q^{1/4}
 a^{-3/4},
\end{equation}
where $ q= \frac{\lambda^2 \phi^2}{m_{\phi}^2}$. This can be used to
calculate the neutrino density
\begin{equation}
\rho_{\nu} \simeq \int_0^{\kappa_F} d^3k = 8 \pi \lambda^2
\frac{1}{2} m_{\phi}^2 {\phi}^2 =  8 \pi \lambda^2 \rho_{\rm DM},
\label{nupro}
\end{equation}
where we have used $\rho_{\rm DM}=\frac{1}{2}m_\phi^2\phi^2$. It is
important to note that the neutrino energy density is proportional
to the local dark matter density. Using this fact and the continuity
equation for the total dark matter and neutrino fluid, we can find
the evolution of the dark matter energy density and hence neutrino
energy density. The continuity equation for the dark matter and
neutrino as a whole reads
\begin{equation}
 \dot{\rho}_{\rm tot}+3H\rho_{\rm tot}\left(1+w_{\rm tot}\right)=0,
 \label{eq:continuity}
\end{equation}
where $w_{\rm tot}=\frac{P_{\rm tot}}{\rho_{\rm tot}}$, $\rho_{\rm
tot}=\rho_\nu+\rho_{\rm DM}$ and $P_{\rm tot}=P_\nu+P_{\rm DM}$.
Eq.~\ref{eq:continuity} can be split up into the two components
\begin{equation}
 \dot{\rho}_{\rm DM}+3H\rho_{\rm DM}=-Q
 \label{eq:contDM}
\end{equation}
and
\begin{equation}
 \dot{\rho}_\nu+4H\rho_\nu=Q,
 \label{eq:contnu}
\end{equation}
where Q represents the decay rate from dark matter to neutrinos and
we have taken advantage of the fact that $P_{\rm DM}=0$ and
$P_\nu=\frac{1}{3}\rho_\nu$. Combining Eqs.~\ref{nupro},
\ref{eq:contDM}, and \ref{eq:contnu} we get the relations
\begin{align}
 \rho_{\rm DM}&=\rho_{\rm DM}^i\left(\frac{a}{a^i}\right)^{-\iota}\nonumber\\
 \rho_\nu&=8\pi \lambda^2\rho_{\rm DM},
\end{align}
where the $i$ denotes the value at some fixed time(e.g. today) and
$\iota=\frac{3+32\pi \lambda^2}{1+8\pi \lambda^2}$. So here we
clearly see that due to parametric production, dark matter no longer
redshifts as $1/a^3$. Its effective equation of state changes from
zero to slightly higher values. The higher the coupling $\lambda$,
the higher the deviation. This effective equation of state which
corresponds to the value we would get if we did not know about the
coupling between dark matter and neutrinos can be calculated from a
revised version of Eq.~\ref{eq:contDM} $\dot{\rho}_{\rm
DM}+3H\rho_{\rm DM}(1+w_{\rm eff})=0$. The result is
\begin{equation}
 w_{\rm eff}=\frac{\iota}{3}-1.
\end{equation}
We note that for $\lambda \rightarrow 0 $, it gives the right limit
for the equation of state $w_{\rm eff} \rightarrow 0$.

\subsection{ Regime II: Fermi radius stops expanding ($q \simeq 1$)}

In the second regime, parametric excitations weaken due to the drop
in resonance parameter $q$. During this regime, $q\sim1$ and the
radius of the physical Fermi sphere has approached the constant
value $k_F\sim m_\phi$. This means the decay is controlled by the
redshifts of Fermi momentum due to the expansion. As the universe
expands, the Fermi momentum drops, opening up
space in the Fermi sphere. This space is immediately filled up by
the scalar field decaying into neutrinos. The regime may be
important for structure formation if the decay into neutrinos can
cause a substantial decrease in the dark matter density. This is
possible when $\rho_{DM} \sim m_{\phi}^4 $, because, in this case,
decay of each DM particle to a neutrino causes a significant
decrease in the dark matter energy density. As the dark matter mass
is of the order sub eV in our model, this can happen only at late
times. This can enhance the late ISW effect thus modifying the
structure formation on large scales. Again, using Eq.~\ref{nupro} we
easily obtain $\rho_\nu=4\pi m_\phi^4$. From the continuity
equations Eq.~\ref{eq:contDM} and Eq.~\ref{eq:contnu} we get the
relations
\begin{align}
 \rho_{\rm DM}&=-\frac{\gamma}{3}+\left(\rho_{\rm
 DM}^i+\frac{\gamma}{3}\right)\left(\frac{a^i}{a}\right)^3\nonumber\\
 \rho_\nu&=\frac{\gamma}{4},
\end{align}
where $\gamma=16\pi m_\phi^4$. So we see that the neutrino energy
density is constant in this regime, only the dark matter density
dilutes. This is what we expect, because if there were no decay into
neutrinos, its density would simply redshift like the standard model
neutrino governed by the Hubble expansion. But here, as soon as
phase space opens in the Fermi sea of neutrinos due to cooling of
the universe, it gets refilled by the decay from dark matter thus
keeping its density constant.

\begin{figure}[!htb]
 \begin{center}
 \includegraphics[width=2.5in]{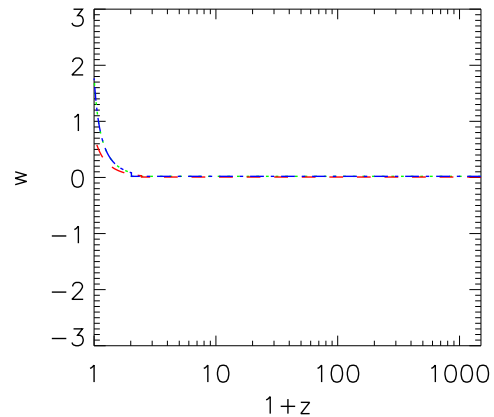}
 \caption{The effective equation of state of dark matter in the preheating dark matter scenario with the following parameter choices:
 Dotted green: $\Omega_{\rm DM,0}=0.21$,
 $\lambda=0.001$ with $\Omega_{\nu,0}=0.022$, dot-dashed blue: $\Omega_{\rm DM,0}=0.06$,
 $\lambda=0.05$ with $\Omega_{\nu,0}=0.081$, and dashed red: $\Omega_{\rm DM,0}=0.12$,
 $\lambda=0.03$ with $\Omega_{\nu,0}=0.068$.}
 \label{fig:wdm}
 \end{center}
\end{figure}

In this regime the effective equation of state for dark matter can
be calculated to be
\begin{equation}
 w_{\rm eff}=\frac{\frac{\gamma}{3}}{\rho_{\rm
 DM}},
 \end{equation}
which depending on model parameters can deviate an appreciable
amount toward the present - see Fig.~\ref{fig:wdm}.

\subsection{Transition redshift}

In this subsection we find the transition redshift between the two
epochs in terms of model parameters $m_{\phi}$ and $\lambda$. Patching the
two regimes together at a scale factor $a_T$ we arrive at the
relations
\begin{equation}
  \begin{array}{l l}
    \begin{array}{l l}
    \rho_\nu =8\pi\lambda^2\rho_{\rm DM} & \quad \mbox{}\\
    \rho_{\rm DM}=\frac{m_\phi^4}{2\lambda^2}\left(\frac{a}{a_T}\right)^{-\iota} & \quad \mbox{}\\
    \end{array} & \quad \mbox{for $a<a_T$}\\
    \\
    \begin{array}{l l}
    \rho_\nu =\frac{\gamma}{4} & \quad \mbox{}\\
    \rho_{\rm DM}=-\frac{\gamma}{3}+\left(\rho_{\rm DM}^0+\frac{\gamma}{3}\right)\left(\frac{a_0}{a}\right)^3, & \quad \mbox{}\\
    \end{array} & \quad \mbox{for $a>a_T$}\\
  \end{array}
  \label{eq:rhos1}
\end{equation}
where the $0$ denotes present day values, and $a_T$ can be
determined from the preheating dark matter parameters as
\begin{equation}
 a_T=\left[\frac{\left(\rho_{\rm
 DM}^0+\frac{\gamma}{3}\right)}{\frac{m_\phi^4}{2\lambda^2}+\frac{\gamma}{3}}\right]^{1/3}a_0.
 \label{eq:atrans}
\end{equation}

%%%%%%%%%%%%%%%%%%%%%%%%%%%%%%%%%%%%%%%%%%%%%%%%%%%%%%%%%%%%%%%%%%%%%%
\section{Perturbation Analysis}%%%%%%%%%%%%%%%%%%%%%%%%%%%%%%%%%%%%%%%
%%%%%%%%%%%%%%%%%%%%%%%%%%%%%%%%%%%%%%%%%%%%%%%%%%%%%%%%%%%%%%%%%%%%%%
In order to study the implications of preheating dark matter, we perform a
cosmological perturbation analysis in the synchronous gauge in which
a line element is given by
\begin{equation}
 ds^2=a^2(\tau)\left[-d\tau^2+(\eta_{ij}+h_{ij})dx^idx^j\right],
\end{equation}
where $i,j=1,2,3$, $\eta_{ij}$ is the Minkowski space metric,
$h_{ij}$ is the perturbation to the metric and we are using comoving
coordinates $x^\mu=(\vec{x},\tau)$ in a spatially flat background
space-time. We follow the procedure given by Ref.~\cite{Ma:1995ey},
in which, to linear order in the perturbations, the stress-energy
tensor is given by
\begin{align}
 T^0_0&=-(\bar{\rho}+\delta\rho)\nonumber\\
 T^0_i&=(\bar{\rho}+\bar{P})v_i=-T^i_0\nonumber\\
 T^i_j&=(\bar{P}+\delta
 P)\delta^i_j+\Sigma^i_j,\,\,\,\,\Sigma^i_i=0,
 \label{eq:stress}
\end{align}
where the perturbations to energy density and pressure are defined
as $\delta\rho=\rho-\bar{\rho}$ and $\delta P=P-\bar{P}$,
$\Sigma^i_j$ is the anisotropic shear perturbation, and $v_i$ is the
coordinate velocity of the fluid\footnote{For further information
about cosmological perturbation theory see Ref.\cite{Kodama:1985bj}}. The latter is
a small quantity and can be treated as a perturbation of the same
order as $\delta\rho$ and $\delta P$. Instead of working with the
velocity itself we use the divergence defined as $\theta=ik^iv_i$.
Similarly, instead of the anisotropic shear perturbation, we use the
shear stress $\sigma$. This is defined as
$\sigma=\frac{-\left(k_ik^j-\frac{1}{3}\eta_{i}^{j}\right)\Sigma^i_j}{\left(\bar{\rho}+\bar{P}\right)}$.

The conservation of energy and momentum for our coupled fluid
implies that the covariant derivative of the stress-energy tensor is
0.
\begin{equation}
\Gamma^{\mu\sigma}_{;\mu\,\,\rm fluid}=\partial_\mu
T^{\mu\sigma}+\Gamma^\sigma_{\alpha\beta}T^{\alpha\beta}+\Gamma^\alpha_{\alpha\beta}T^{\sigma\beta}=0.
\end{equation}
However, for the individual components in the fluid it is slightly
different
\begin{align}
\Gamma^{\mu\sigma}_{;\mu\,\,\rm CDM}&=-\delta Q\nonumber\\
\Gamma^{\mu\sigma}_{;\mu\,\,\rm \nu}&=\delta Q, \label{eq:dq}
\end{align}
where the $\delta Q$ can be determined directly from
Eq.~\ref{eq:contDM}. Using the time-time(00) component of the
stress-energy tensor from Eq.~\ref{eq:stress} and combining with
Eq.~\ref{eq:dq} above, we get the equation of motion for the
individual density contrasts $\delta_i=\frac{\delta\rho_i}{\rho_i}$.
Using the space-space(ii) components in the same way will give us
the time evolution of $\theta_i$.

Hence we arrive at the equations of motion for the DM and neutrino
components for $q\gg1$
\begin{align}
 &\dot{\delta}_\nu=-\frac{4}{3}\left(\theta_\nu+\frac{\dot{h}}{2}\right)\nonumber\\
 &\dot{\delta}_{\rm CDM}=-\frac{\dot{h}}{2}\nonumber\\
 &\dot{\theta}_\nu=-\frac{1}{1+8\pi\lambda^2}H\theta_\nu+k^2\left(\frac{1}{4}\delta_\nu-\sigma_\nu\right)\nonumber\\
 &\dot{\theta}_{\rm CDM}=0.
\end{align}
Similarly, for $q\sim1$ we get
\begin{align}
 &\dot{\delta}_\nu=-\frac{4}{3}\left(\theta_\nu+\frac{\dot{h}}{2}\right)\nonumber\\
 &\dot{\delta}_{\rm CDM}=-\frac{\dot{h}}{2}+4H\frac{\rho_\nu}{\rho_{\rm CDM}}\left(\delta_{\rm CDM}-\delta_\nu\right)\nonumber\\
 &\dot{\theta}_\nu=-4H\theta_\nu+k^2\left(\frac{1}{4}\delta_\nu-\sigma_\nu\right)\nonumber\\
 &\dot{\theta}_{\rm CDM}=0.
 \label{eq:deltatheta2}
\end{align}

It is the effect of the term containing $\lambda$ and the term
containing the neutrino energy density $\rho_\nu$ in the evolution
of the CDM density contrast that separates the evolution of
perturbations in the preheating dark matter case from the normal case, where the decay
of CDM is not permitted.

%%%%%%%%%%%%%%%%%%%%%%%%%%%%%%%%%%%%%%%%%%%%%%%%%%%%%%%%%%%%%%%%%%%%%%
\section{Results}%%%%%%%%%%%%%%%%%%%%%%%%%%%%%%%%%%%%%%%%%%%%%%%%%%%%%
%%%%%%%%%%%%%%%%%%%%%%%%%%%%%%%%%%%%%%%%%%%%%%%%%%%%%%%%%%%%%%%%%%%%%%

In this section we present the numerical results we obtained by
using the equations from the two previous sections. We modified the
publicly available CMBFAST code \cite{CMBFAST} to include preheating dark matter to
get both the matter power spectrum and the temperature anisotropy
spectrum. This code is developed to calculate the linear CMB
anisotropy spectra based on integration over the sources along the
photon past light cone, but also outputs transfer functions from
which the linear matter power spectrum can be calculated.

In our analysis we keep the epoch of matter-radiation equality
fixed, and the only free parameters are the current value of
$\Omega_{\rm DM,0}$ and the parameter $\lambda$. In addition, we
keep the amount of baryons today fixed at $\Omega_b=0.05$ and choose
the normalized Hubble expansion rate at the value $h_{\rm
Hubble}=0.7$. We include one species of massless neutrinos produced
in the decay as well as the three standard model neutrinos, which
for simplicity are assumed to have a degenerate mass spectrum with
$m_\nu=1.5\times10^{-3}$ eV. We assume that the neutrinos produced in the decay
mix with standard neutrinos, although this assumption makes no
qualitative difference to the results.

\begin{figure}[!htb]
 \begin{center}
 \includegraphics[width=2.5in]{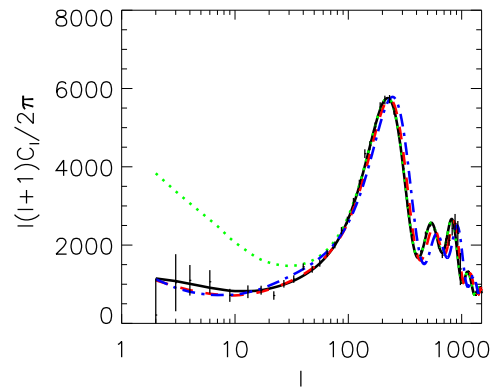}
 \caption{The temperature anisotropy spectrum as a function of the angular modes in the preheating dark matter scenario with the following parameter choices:
 Solid black: $\Lambda$CDM with $\Omega_{\rm DM,0}=0.24$,
 $\Omega_{\Lambda,0}=0.71$, dotted green: $\Omega_{\rm DM,0}=0.21$,
 $\lambda=0.001$ with $\Omega_{\nu,0}=0.022$, dot-dashed blue: $\Omega_{\rm DM,0}=0.06$,
 $\lambda=0.05$ with $\Omega_{\nu,0}=0.081$, and dashed red: $\Omega_{\rm DM,0}=0.12$,
 $\lambda=0.03$ with $\Omega_{\nu,0}=0.068$.}
 \label{fig:ani}
 \end{center}
\end{figure}

The impact of preheating dark matter on the temperature anisotropy
spectrum, can be seen in Fig.~\ref{fig:ani}. The most apparent
difference from the spectrum of $\Lambda$CDM can be seen on largest
angular scales, $l\lesssim100$ (corresponding roughly to a degree),
although for some choices of parameters the positions and relative
heights of the peaks are also affected. We generally observe an
increase in power on scales, $20<l<100$, whereas on scales
$l\lesssim10$ we see an increase or a decrease in power depending on
the model parameters. For scales $l\lesssim100$, the dominant
contribution to the CMB temperature anisotropy spectrum comes from
the Integrated Sachs-Wolfe effect, which arises because of the
evolution of the gravitational potentials encountered by the
photon on its journey from the last-scattering surface. The
modification to the cosmological background because of preheating dark matter can be
quite significant, as we saw in the last section. In particular in
the $q\sim1$ regime, where the neutrino energy density becomes
constant, which leads to a second term in the evolution of the CDM
perturbation in Eq.~\ref{eq:deltatheta2}.

On the largest angular scales $l<20$, the dominant contribution to
the CMB temperature anisotropy spectrum comes from the late-time
Integrated Sachs-Wolfe effect (ISW). This effect is a result of the
universe entering an epoch of rapid expansion as it becomes
dominated by dark energy. In this epoch, photons moving into
gravitational potential wells will get a boost as the potential well
is decaying and becomes slightly shallower while the photon is
passing through it - and vice versa for gravitational hills. It is
clear that this effect depends intimately on many different
parameters such as $\Omega_{\rm DM,0}$ and $\Omega_{\nu,0}$.
In our model, dark matter is being transformed into light neutrinos - i.e.
radiation - most efficiently at late times, and hence we expect an
effect on the largest scales. In the context of interacting dark matter-dark energy models,
the ISW effect has been studied in \cite{Olivares:2008bx}.

\begin{figure}[!htb]
 \begin{center}
 \includegraphics[width=2.5in]{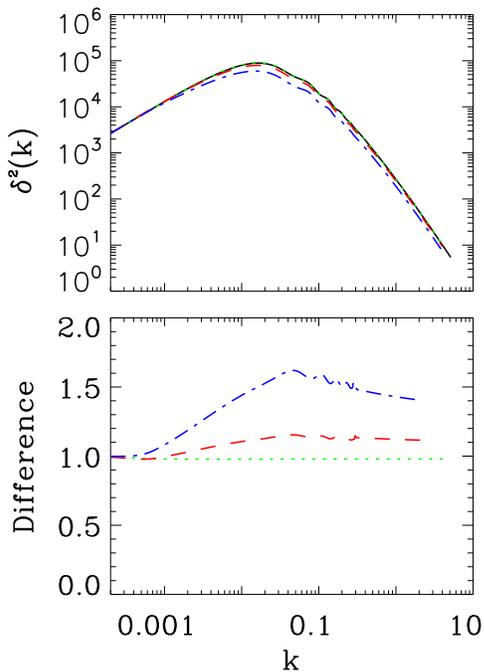}
 \caption{Top: The matter power spectra as a function of k[Mpc$^{-1}$] in the preheating dark matter scenario with the following parameter
 choices: Solid black: $\Lambda$CDM with $\Omega_{\rm DM,0}=0.24$,
 $\Omega_{\Lambda,0}=0.71$, dotted green: $\Omega_{\rm DM,0}=0.21$,
 $\lambda=0.001$ with $\Omega_{\nu,0}=0.022$, dot-dashed blue: $\Omega_{\rm DM,0}=0.06$,
 $\lambda=0.05$ with $\Omega_{\nu,0}=0.081$, and dashed red: $\Omega_{\rm DM,0}=0.12$,
 $\lambda=0.03$ with $\Omega_{\nu,0}=0.068$.
 Bottom: The difference between the preheating dark matter model and $\Lambda$CDM
 ($\delta^2$($\Lambda$CDM)/$\delta^2$(PHDM)) as a function of k.}
 \label{fig:delta}
 \end{center}
\end{figure}

Turning our attention to the matter power spectra, we know that the
linear matter power spectrum is extremely well determined by SDSS
(see Ref.~\cite{Tegmark:2003ud}) and WMAP (see
Ref.~\cite{Komatsu:2008hk}) on intermediate and large scales. And in
addition, Ly-$\alpha$ forest data have some constraints on the small
scales - see e.g. \cite{Zaroubi:2005xx}. Hence we normalize our
matter power spectra such that they coincide with matter power
spectra obtained from using normal $\Lambda$CDM at the largest
scales (these are also the latest to have entered the horizon). The
results are presented in Fig.~\ref{fig:delta} for different values
of $\lambda$ and $\Omega_{\rm DM,0}$. A small damping on small
angular scales seems to be generic, similar to standard models of
CDM and hot dark matter, where a similar reduction in power is
achieved. We note that in order to comply with e.g. supernova data
(Ref.~\cite{Kowalski:2008ez}) we cannot change $\Omega_{\rm M,0}$
drastically. Of course we do turn CDM into neutrinos that redshift
as radiation - only slightly faster than CDM. Hence we have some
room to change $\Omega_{\rm M,0}$ and still be in agreement with data.

The matter power spectra for the different parameter choices
are agreeing relatively well with $\Lambda$CDM as result of the
small $\lambda$-value. Still, we do notice the small reduction on the
smallest scales which can be probed by CMBFAST. This is to be
expected since part of the CDM responsible for the gravitational
wells is decaying into neutrinos which undergo free-streaming on
these small scales. This prevents them from clustering and since we
are creating an appreciable amount of neutrinos due to the decay of
CDM, we generally expect this reduction of power on small scales.

On even smaller scales we also expect a considerable effect, which
will reduce the clustering ability of CDM. Unfortunately, we cannot
probe those scales satisfactorily with CMBFAST - as it is using
linear perturbation analysis and we expect the evolution to be
highly non-linear. The investigation of the effect of preheating dark matter on
non-linear scales is beyond the scope of the present work and will
be postponed to a future paper.

%%%%%%%%%%%%%%%%%%%%%%%%%%%%%%%%%%%%%%%%%%%%%%%%%%%%%%%%%%%%%%%%%%%%%%
\section{Conclusion} %%%%%%%%%%%%%%%%%%%%%%%%%%%%%%%%%%%%%%%%%%%%%%%%
%%%%%%%%%%%%%%%%%%%%%%%%%%%%%%%%%%%%%%%%%%%%%%%%%%%%%%%%%%%%%%%%%%%%%%
In this work, we have considered a coherently oscillating scalar
field of sub-eV mass which behaves as dark matter. Due to its
coupling, it slowly decays into neutrinos as the universe expands
until today. The decay rate as a function of redshift has been
derived following the physics of inflationary preheating of a
scalar into fermions. We find that the decay rate is modulated by
Pauli blocking and the expansion of the universe, giving us rich physics
of dark matter decay into a neutrino sea.

We studied the effect of the decay on structure formation and
obtained spectra for the anisotropies in the cosmic microwave
background and matter power spectra. For the parameters proposed in
this paper, we showed that given the decay we are able to slightly
reduce the amount of power on small scales in the matter power
spectra - something which seems to be required from data - while at
the same time being in good agreement with SDSS and WMAP
observations.

Interestingly, the proposed scenario leads to features in the
temperature anisotropy spectrum - which can be seen as a prominent
late ISW effect and slight modifications to the second and third
peaks. Consequently, as future direction, it would be interesting to
study the late ISW effect in detail, from which we will be able to
constrain the model parameters more effectively. In addition, we
would like to do a follow-up COSMOMC analysis using the newest data
available as well as to examine the effect of preheating dark matter
on structure formation on non-linear scales. After such an analysis
it will be more clear what the best choice of preheating dark matter
parameters is such that a comparison with future Planck data, for
instance, will be easier. Furthermore we expect future weak lensing
surveys to be useful in constraining our scenario since they can
provide insight into the statistics of the dark matter distribution
- hence they can (hopefully) shed some light on what happens on
small (and large) scales of gravitational clustering.

%%%%%%%%%%%%%%%%%%%%%%%%%%%%%%%%%%%%%%%%%%%%%%%%%%%%%%%%%%%%%%%%%%%%%%
\section*{Acknowledgements}%%%%%%%%%%%%%%%%%%%%%%%%%%%%%%%%%%%%%%%%%%%
%%%%%%%%%%%%%%%%%%%%%%%%%%%%%%%%%%%%%%%%%%%%%%%%%%%%%%%%%%%%%%%%%%%%%%
We would really like to thank Neal Weiner for ideas and discussions
and motivating this work. OEB thanks Steen Hannestad and Yvonne Wong
for helpful discussions during this work. SD thanks Kris Sigurdson
for some fruitful suggestions. The research of SD is supported by
the Natural Sciences and Engineering Research Council of Canada.

\end{document}